\documentclass[12pt,leqno]{article}
\usepackage{url}
\usepackage{xspace}

\usepackage{amsmath}
\usepackage{amssymb}

\usepackage{pifont}

\let\ensuremathTEMP\ensuremath

\makeatletter%
\def\nottoobig#1{{\hbox{$\left#1\vcenter to1.111\ht\strutbox{}\right.\n@space$}}}
\makeatother%

\makeatother%

\makeatletter%

\newcount\hour  \newcount\minutes  \hour=\time  \divide\hour by 60
\minutes=\hour  \multiply\minutes by -60  \advance\minutes by \time
\def\mmmddyyyy{\ifcase\month\or Jan\or Feb\or Mar\or Apr\or May\or Jun\or Jul\or
  Aug\or Sep\or Oct\or Nov\or Dec\fi \space\number\day, \number\year}
\def\hhmm{\ifnum\hour<10 0\fi\number\hour :%
  \ifnum\minutes<10 0\fi\number\minutes}
\def\Draft{{\it Draft of \mmmddyyyy}}

\def\ps@jtsheadings{%
\def\@oddhead{\it\rightmark\hfil\rm\thepage}%
\def\@oddfoot{\hfil\Draft}%
\if@twoside%
\def\@evenhead{\rm\thepage\hfil\it\leftmark}%
\def\@evenfoot{\Draft\hfil}%
\else
\let\@evenhead\@oddhead%
\let\@evenfoot\@oddfoot%
\fi%
}
\def\ps@jtsplain{%
\def\@oddhead{\hfil\Draft}%
\def\@oddfoot{\hfil\rm\thepage\hfil}%
\let\@evenfoot\@oddfoot%
\if@twoside \def\@evenhead{\Draft\hfil} \else \let\@evenhead\@oddhead \fi
}

\def\chaptermark#1{\markboth{\thechapter.\ #1}{\thechapter.\ #1}}%
\def\sectionmark#1{\markright{\thesection.\ #1}}

\def\section{\@startsection {section}{1}{\z@}
    {3.5ex plus1ex minus.2ex}{2.3ex plus.2ex}{\Large\bf}}
\def\subsection{\@startsection{subsection}{2}{\z@}
    {3.25ex plus1ex minus.2ex}{1.5ex plus.2ex}{\large\bf}}
\def\subsubsection{\@startsection{subsubsection}{3}{\z@}
    {3.25ex plus1ex minus.2ex}{1.5ex plus.2ex}{\normalsize\bf}}
\def\paragraph{\@startsection{paragraph}{4}{\z@}
    {3.25ex plus1ex minus.2ex}{1em}{\normalsize\bf}}
\def\subparagraph{\@startsection{subparagraph}{4}{\parindent}
    {3.25ex plus1ex minus.2ex}{1em}{\normalsize\bf}}

\makeatother%

\makeatletter \@beginparpenalty=10000 \makeatother

\def\underl#1 {\leavevmode\let\first=\relax\underli #1 }
\def\underli#1 {\ifx&#1\let\next=\relax\unskip
                \else\let\next=\underli\first\ulinebox{#1}\fi\let\first=\undersp\next}
\def\undersp{\penalty50\ulinebox{\space}\penalty50}
\def\ulinebox#1{\vtop{\hbox{\strut#1}\hrule}}%
\def\unice#1 {\underl #1 & }
\def\desclabel#1{\bf #1\hfil}
\def\desc{\list{}{%
\setlength{\leftmargin}{0pt}
\labelwidth= \leftmargin
\advance \labelwidth by -\labelsep
\let \makelabel=\desclabel}}

\newcounter{extremeleftlistcounter}
  {\begin{list}{\arabic{extremeleftlistcounter}~~~}{\usecounter{extremeleftlistcounter}%
        \setlength{\labelsep}{0pt}\setlength{\leftmargin}{0pt}%
        \setlength{\labelwidth}{0pt}\setlength{\listparindent}{0pt}}}%
  {\end{list}}

\newcounter{leftlistcounter}
  {\begin{list}{\arabic{leftlistcounter}~~~}{\usecounter{leftlistcounter}%
        \setlength{\labelsep}{0pt}\setlength{\leftmargin}{15pt}%
        \setlength{\labelwidth}{15pt}\setlength{\listparindent}{0pt}}}%
  {\end{list}}



\makeatletter %
\newcommand{\trueit}{\mbox{\rm true}}
\newcommand{\falseit}{\mbox{\rm false}}
\newcommand{\truerm}{\mbox{\rm{}true}}
\newcommand{\falserm}{\mbox{\rm{}false}}

\newlength{\filength}
\newsavebox{\gcbox}
\sbox{\gcbox}{\framebox[\filength]{\rule{0ex}{2ex}}}

\newlength{\leftjustindent}
\newlength{\@leftjustindent}
\def\leftjust{\let\\\@leftjustcr\let\end\@endleftjust
  \addtolength{\@leftjustindent}{\leftjustindent} \vcenter\bgroup
\halign\bgroup \hbox to\displaywidth{
\rule{\@leftjustindent}{0ex}$\displaystyle##$\hfill }\crcr }
\def\endleftjust{\crcr\egroup\egroup\endgroup}
\def\@endleftjust#1{\crcr\egroup\egroup\@checkend{#1}\endgroup}
\def\@leftjustcr{\crcr}

\newcommand{\singlespacing}{\let\CS=
\@currsize\renewcommand{\baselinestretch}{1}\tiny\CS}
\newcommand{\singlespacingplus}{\let\CS=
\@currsize\renewcommand{\baselinestretch}{1.25}\tiny\CS}
\newcommand{\doublespacing}{\let\CS=
\@currsize\renewcommand{\baselinestretch}{1.75}\tiny\CS}
\newcommand{\draftspacing}{\let\CS=
\@currsize\renewcommand{\baselinestretch}{2.0}\tiny\CS}

\makeatother%

\mathcode`\0="0030      %
\mathcode`\1="0031
\mathcode`\2="0032
\mathcode`\3="0033
\mathcode`\4="0034
\mathcode`\5="0035
\mathcode`\6="0036
\mathcode`\7="0037
\mathcode`\8="0038
\mathcode`\9="0039
\makeatletter
\clubpenalty=\@highpenalty
\widowpenalty=\@highpenalty
\makeatother

\makeatletter
\newcommand{\niceonespacing}{\let\CS=\@currsize\renewcommand{\baselinestretch}{1.1}\tiny\CS}\newcommand{\nicetwospacing}{\let\CS=\@currsize\renewcommand{\baselinestretch}{1.2}\tiny\CS}
\newcommand{\nicethreespacing}{\let\CS=\@currsize\renewcommand{\baselinestretch}{1.3}\tiny\CS}
\newcommand{\singlespacingplusplus}{\let\CS=\@currsize\renewcommand{\baselinestretch}{1.35}\tiny\CS}
\newcommand{\nicefourspacing}{\let\CS=\@currsize\renewcommand{\baselinestretch}{1.4}\tiny\CS}
\newcommand{\nicefivespacing}{\let\CS=\@currsize\renewcommand{\baselinestretch}{1.5}\tiny\CS}
\newcommand{\nicesixspacing}{\let\CS=\@currsize\renewcommand{\baselinestretch}{1.6}\tiny\CS}
\makeatother

\makeatletter
\def\@cite#1#2{[#1\if@tempswa , #2\fi]}
\makeatother

\makeatletter
\def\@citex[#1]#2{\if@filesw\immediate\write\@auxout{\string\citation{#2}}\fi
  \def\@citea{}\@cite{\@for\@citeb:=#2\do
    {\@citea\def\@citea{,\linebreak[0]}\@ifundefined
       {b@\@citeb}{{\bf ?}\@warning
       {Citation `\@citeb' on page \thepage \space undefined}}%
\hbox{\csname b@\@citeb\endcsname}}}{#1}}
\makeatother

\makeatletter
\def\ps@thesis{\def\@oddhead{\hfil\rm\thepage\hfil}\def\@oddfoot{}\def\@evenhead{\hfil\rm\thepage\hfil}\def\@evenfoot{}\def\chaptermark##1{}\def\sectionmark##1{}}
\makeatother

\makeatletter
\def\foobarpt{\textfont\z@\tenrm 
  \scriptfont\z@\ninrm \scriptscriptfont\z@\sevrm
\textfont\@ne\tenmi \scriptfont\@ne\ninmi \scriptscriptfont\@ne\sevmi
\textfont\tw@\tensy \scriptfont\tw@\ninsy \scriptscriptfont\tw@\sevsy
\textfont\thr@@\tenex \scriptfont\thr@@\tenex \scriptscriptfont\thr@@\tenex
\def\unboldmath{\everymath{}\everydisplay{}\@nomath\unboldmath
          \textfont\@ne\tenmi 
          \textfont\tw@\tensy \textfont\lyfam\tenly
          \@boldfalse}\@boldfalse
\def\boldmath{\@ifundefined{tenmib}{\global\font\tenmib\@mbi\@magscale1\global
        \font\tensyb\@mbsy \@magscale1\global\font
         \tenlyb\@lasyb\@magscale1\relax\@addfontinfo\@xiipt
              {\def\boldmath{\everymath
                {\mit}\everydisplay{\mit}\@prtct\@nomathbold
                \textfont\@ne\tenmib \textfont\tw@\tensyb 
                \textfont\lyfam\tenlyb\@prtct\@boldtrue}}}{}\@xiipt\boldmath}%
\def\prm{\fam\z@\tenrm}%
\def\pit{\fam\itfam\tenit}\textfont\itfam\tenit \scriptfont\itfam\ninit
   \scriptscriptfont\itfam\sevit
\def\psl{\fam\slfam\tensl}\textfont\slfam\tensl 
     \scriptfont\slfam\tensl \scriptscriptfont\slfam\tensl
\def\pbf{\fam\bffam\tenbf}\textfont\bffam\tenbf 
   \scriptfont\bffam\ninbf \scriptscriptfont\bffam\ninbf 
\def\ptt{\fam\ttfam\tentt}\textfont\ttfam\tentt
   \scriptfont\ttfam\nintt \scriptscriptfont\ttfam\nintt 
\def\psf{\fam\sffam\tensf}\textfont\sffam\tensf
    \scriptfont\sffam\tensf \scriptscriptfont\sffam\tensf
\def\psc{\@getfont\psc\scfam\@xiipt{\@mcsc\@magscale1}}%
\def\ly{\fam\lyfam\tenly}\textfont\lyfam\tenly 
   \scriptfont\lyfam\ninly \scriptscriptfont\lyfam\sevly
 \@setstrut \rm}

\makeatother

\newcommand{\sat}{{\rm SAT}}
\newcommand{\satbar}{{\overline{\rm SAT}}}

\newcommand{\p}{{\rm P}}

\newcommand{\np}{{\rm NP}}

\singlespacingplus

\newcommand{\condition}{\,\ensuremathTEMP{\mbox{\large$|$}}\:}

\def\land{{\; \wedge \;}}

\newenvironment{block}{\begin{list}{\hbox{}}{\leftmargin 1em
    \itemindent -1em \topsep 0pt \itemsep 0pt \partopsep 0pt}}{\end{list}}

\makeatletter
\def\@listI{\leftmargin\leftmargini \parsep 4.5pt plus 1pt minus 1pt\topsep
6pt plus 2pt minus 2pt \itemsep  2pt plus 2pt minus 1pt}

\let\@listi\@listI
\@listi
\makeatother

\makeatletter %
 \newcommand{\setoffdisplay}{\rule{5.9in}{1pt}}

\makeatother



\usepackage{fullpage}
\usepackage{graphicx}
\usepackage{color}
\usepackage{parskip}  %
\usepackage{amssymb}
\usepackage[absolute]{textpos} %
\usepackage{tikzit}

\usepackage{fancyhdr}
\fancypagestyle{plain}{%
    \fancyhf{}%
\fancyfoot[C]{Introduction Page \#\thepage}%
}

\begin{document}
\singlespacing

\title{Fortune's Bounty: Taming Complexity by Trimming Trees~---~A Hands-On Problem-Solving Experience in Advanced
  Complexity Suitable for Introductory Students\thanks{Supported in part by NSF grants CCF-2006496, DUE-2135431, and DUE-213543, and a Renewed Research Stay grant from the Alexander von Humboldt Foundation.}}

\author{Kimberly Fluet\\Warner School of Education\\University of Rochester\\Rochester, NY, USA\\
  {\tt{}kfluet@warner.rochester.edu}
  \and
Lane A. Hemaspaandra\thanks{Work done in part while on sabbatical at the University of D\"{u}sseldorf.}\\Department of Computer Science\\University of Rochester\\Rochester, NY, USA\\\url{cs.rochester.edu/~lane}
\and
Christopher M. Homan\\
Department of Computer Science\\
Rochester Institute of Technology\\Rochester, NY, USA\\{\tt{}cmh@cs.rit.edu}}
\date{August 13, 2026}

\maketitle

\begin{abstract}
  This article provides an assignment designed to let undergraduate students who have completed an undergraduate CS1/CS2 sequence try to themselves, in groups, prove Fortune's Theorem. (Fortune's Theorem states that if the complement of the Boolean satisfiability problem polynomial-time reduces to a sparse set, then the Boolean satisfiability problem is polynomial-time computable.  The assignment does not assume that students have previously seen the Boolean satisfiability problem, polynomial-time reductions, or sparse sets. Rather, it teaches those within the assignment.  Note: Reworded into the technical vocabulary of complexity theory, Fortune's Theorem states that no sparse set is coNP-hard unless $\p=\np$.  Fortune's Theorem was a major advance in the understanding of the relationship between hardness and density.)

  We provide both the assignment handout (as the main body of this report plus Appendix A) and a solution to the assignment (as Appendix B, which would of course not be made available to the students until after they had handed in the assignment). The assignment handout, though the instructor can change this, is framed as having the students starting the assignment in teams in class for a whole class session, and then finishing it in those same teams as a take-home assignment, and handing it in before the next class session.

  We have found that student groups often succeed, partially or completely, in this challenge. This can mean a lot to the students: they see that they were able to make an advance that, when it was first obtained, appeared in what was arguably at the time the top journal venue for complexity theory research. This can give them confidence that they have substantial problem-solving skills (which basically means research skills) when they truly apply themselves to a given challenge.
\end{abstract}

\section{Introduction}

\subsection{Pr\'{e}cis}
The main goal of this assignment is to give students confidence that they can do research. The assignment does so by putting them, as part of a team, in the same position that Steve Fortune was in, in the 1970s, when he proved--and published in a top CS theory journal--the historically important result that the complement of SAT (where SAT is the NP-complete Boolean satisfiability problem) cannot mapping (aka many-one) reduce to a sparse set unless P=NP (A Note on Sparse Complete Sets, Steven Fortune, SIAM Journal on Computing, 1979). That is, students see that they too can make that same advance as Fortune, given no more (and in fact slightly less) groundwork than Fortune himself had from earlier papers.

The material that the students will be exploring as they tackle this is about, under the abovementioned hypothesis about a reduction to a sparse set existing, finding good rules to on-the-fly prune potentially exponential-sized trees in such a way that they become polynomial in size.

The assignment does this in such a way that students, while they do the assignment, do not have to know or learn what a nondeterministic or co-nondeterministic Turing machine is, or what the complexity classes NP and coNP are. The assignment can be used to show students, even quite early in their careers, that they can achieve what was, in its time, a major cutting-edge advance.

\subsection{Logistics, Background, and Desired Benefits}
The assignment a professor would---after re-editing the deadline, collection mechanism, and other logistic details to their course's needs, see the next paragraph---hand out to students would be all the pages of this document from the one right after the current page through and including Appendix A. We have tweaked the page numbering so that the next page becomes page 1 in its numbering, and so on, each with ``Assignment'' mentioned in its page-numbering footer, so that those pages can be easily identified, lifted out, and given to students.

Of course, those pages in their ``Rules'' have some items, such as the deadline and where/how to turn the item in, that the teacher would need to edit or fill in. We will happily share with any professor who writes to us from their institutional email address the source code of this article, so that they can edit the rules/etc.~to fit their own dates, hand-in mechanism, and rules (and can even, if they wish, change the solutions to different solutions).

In Appendix B, we provide sample solutions to the assignment's questions.  Those, of course, would not be made available to students until after they had completed and handed in their work on the assignment.

Please do not worry about the fact that the bibliography/references are included in the solutions, and so will not be handed to students when they are given the assignment pages. The assignment pages, on purpose, include no ``$\backslash$cite'' calls, and so never draw on the bibliography. Rather, the intent is that the students will get the bibliography and the solution's literature discussion when they, after tackling and submitting the assignment's challenge, are given access to the solutions.

We have used this assignment in various settings. Most often, we give it not as a start-in-class in teams and complete in those same  teams as a take-home assignment, which is how it is framed here, but as an in-class, full-class-session, team-based problem-solving challenge. However, the time limits of that setting increase the already substantial level of challenge to the students, and so this document frames the assignment in a way that gives students more time to reach insights and success. 
In the article's abstract, we stated that the assignment can be tackled by students whose background is just CS1+CS2. That is true, but we in fact have given it in classes where the students are not all CS majors, and thus some of the class's students have had neither CS1 nor CS2.  In such settings, we carefully set up the teams so that students with no CS background have as teammates students with a CS background.  The courses we have given this hands-on experience in range from undergraduate problem-solving seminars to a ``gems'' of theoretical computer science course that does not require prior CS experience, to an undergraduate introduction to models of computation course, to a PhD-level complexity theory course.

What is consistent over all these varied uses is that students seem to truly value and enjoy immersing themselves in problem-solving. For a class session---or if used as a take-home assignment or a start-in-class-but-finish-as-take-home assignment, for the days of the assignment---they are not being lectured at in class, and they are not being given a ``routine'' assignment, but rather they are themselves seeking to make, on a level playing field with what was historically the case, a major advance in theoretical computer science (albeit, one that was made in the past). This experience not only introduces students to the joy of team-based research, but while doing so also helps the students realize that they too can make advances.

\clearpage
\setcounter{page}{1}
\fancyfoot[C]{Assignment Page \#\thepage}

\sloppy
\begin{center}
    {\Large
      Fortune's Bounty:
      Taming Complexity by Trimming Trees}
\end{center}

{\em Rules:

  You will do this assignment in the groups of size about 4 (each group has either 4 or 5 members) that are projected on the screen and also are in our Blackboard course area's Groups folder.
 
  Your group will have this entire class session to start this assignment, and then can continue to work on it until the hand-in deadline.  Your group's answers must be handed in no later than five minutes before the start of our next class session, so you'll have until shortly before Monday's class session---about five days---to continue working on this.  During this class session, the TA and I will circulate as you work on this; we will not give you answers, nor will we give you feedback on answers you have, but if you have questions about the definitions/notions/etc., we will be happy to answer those questions. Please do make sure that by the end of this class session you understand well what the assignment is asking---the definitions, notations, and framework, and the nature of the expected turn-in.

  Each group should hand in its answers by having one group member email a PDF file---containing the full names of all members of your group and your group's answer for each of this assignment's two questions---to our course's standard assignment-turn-in email address, \mbox{\tt{}xxxx@xxxx.xxxx.xxxx}. Your group member who sends that email should carbon-copy each group member, so that all the group's members know that the group's turn-in has been submitted. (Your group's turn-in should not give separate answers, one from each group member, but rather should give one answer to each of the two questions, namely, the
  the answer your group converged on.)

  We will go over the answers to this assignment at the start of our next class (likely inviting groups to present their answers if they wish), and also, shortly after our next class, I'll post written answers in our course's Blackboard area. Additionally, each group will receive feedback from me and/or the TA on its turn-in.

  But the feedback will not be a grade. This is not a graded item. I want you to focus on working with your teammates on the (substantial) challenge here, not on trying to optimize a grade. I also want to make sure that you don't (as having this be for a grade would incentivize) either lunge for the hint section before your group has had a chance to think about the challenge without the hint, or during the days between this class and the next try looking up the answers on the Internet or in the library. (After all, this is not a course on web-searching for answers. Rather, the assignment's goal, in addition to its technical content, is to build your problem-solving---and in particular your group problem-solving---skills (and with luck to help you realize that those skills are quite strong). Also, even if your group does not solve both the problems, you will far more deeply appreciate the solutions if your group has worked hard on the problems, thus learning what the obstacles are to a solution and then seeing how the solution overcomes those.)

  The problem is challenging, but I have confidence you can do some or all of it, as many other groups have done in past years. My best advice to you is: Use this initial class session to, with your groupmates, quick-read the assignment and learn the problem's notions---getting help on that from me or the TA by asking us questions during this session---so that your group understands what the notions, framework, and goal are; work hard and meet multiple times with your groupmates in the five days after this class that you have to work on solving the problem as a group, listen to your groupmates' ideas, and share with them your ideas---after all, most strong research is done by teams; and try to enjoy both the teamwork and the challenge of the problem.
 
  As to what resources you can access (even during the out-of-class days), this assignment is ``closed everything'' other than your
  and your groupmates' minds.
}

\subparagraph*{Introduction: Your Group Is Being Challenged to Make an Advance That Was First Achieved in an Important Journal Paper by Steve Fortune}
Your group's task is to prove an important theorem originally proven by Steve Fortune in 1979.  At the time, the top two journals for CS-theory results were Journal of the ACM (JACM) and SIAM Journal on Computing (SICOMP), and Fortune's paper appeared in SICOMP\@.  It will take a while for us to present the concepts needed to so that you can understand what Fortune's theorem says; but just so you can see where we're going (but don't worry if this at this point makes no sense to you as it is mostly jargon and notation), what Fortune's theorem states is that if the complement of the Boolean satisfiability problem polynomial-time reduces to a sparse set, then the Boolean satisfiability problem is polynomial-time computable.  (Yes, it is an intimidating theorem statement.  But you'll soon understand the theorem statement, and with luck, during this assignment your group will itself prove Fortune's theorem via developing powerful tree-pruning rules!)

Fortune's proof built on a framework that had already been developed. I'll give you most of that framework soon, except I'll leave out (as it will be Question~1 here) one important insight that Fortune was lifting from an important earlier paper in this line of work. So you'll have no more prior-framework-to-build-on in hand than Fortune himself had---in fact, you'll have a bit less (unless you look at the Hint appendix to see the hint I give for Question~2).

Fortune added to the existing, standard framework a
simple-but-lovely insight (our Question~2 issue), in a way that (along with being bright enough to use also the insight that Question~1 is about) let him achieve his important result; he rightfully was much acclaimed for doing so, but also his work was a generous contribution (one might say a bounty) to the field, and it in time led to a line of even more powerful results.

Yet in truth, Fortune's lovely insight is something that fifth-graders on a playground would, if presented with it framed in real-world terms (as it is in the Hint section), nod and say ``Duh!''  Of course, the key point here is that from the thousands of things one might try, Fortune found one that actually works.  Your challenge is to do the same.

You might or might not succeed in solving Question~1 and/or Question~2, though I suspect you will solve Question~1 and may well solve Question~2.  Even if your group solves neither of the two questions, I suspect that after seeing the solutions you'll realize that you were close to solving at least one of the questions.  In fact, a key take-away from this assignment---one far more important to your career arc than the technical result itself will be---is that major advances in theoretical computer science are usually not made by deities who have nothing in common with you, but rather are made by people just like you, who look at a challenge, and persistently and flexibly try a variety of approaches until they find a way to solve the challenge.

\subparagraph{Preliminaries/Definitions/Background/Framework}

I assume that you know what a propositional (i.e., quantifier-free) Boolean formula is, what a polynomial-time algorithm is, and what a tree is, and that you understand what it means to say a set (i.e., language) is computable in polynomial time and what it means to say a function is computable in polynomial time. Aside from that, the assignment is self-contained. It does not assume that you know, for example, what NP or coNP are, or what a nondeterministic or co-nondeterministic Turing machine is (though under its hood, this assignment is having you prove a major result that usually assumes you know about such things).

Recall that a Boolean formula is built on Boolean (i.e., taking on the values False or True) variables (e.g., $x_1$), using logical operations.  For this assignment, we will assume (note: in the context of this assignment, that assumption can be made without loss of generality) that the only logical operators allowed in our formulas are the binary operators $\land$ (AND) and $\lor$ (OR), and the unary operator $\neg$ (negation). As is commonly done, we will denote the negation of a variable $x_1$ not by $\neg x_1$ but by $\overline{x_1}$.  We use $F(x_1,x_2, \dots,x_m)$ to denote a Boolean formula over $m$ variables (and for simplicity, throughout this exercise, we assume the variable names in each input formula with $m$ variables are $x_1,x_2, \dots,x_m$).
For example, we might have $F(x_1,x_2, x_3,x_4) = (x_1 \land \overline{x_3}) \lor (x_2 \land \neg (x_3 \land x_1))$.

We say a Boolean formula $F$ is satisfiable if there is some assignment to its variables under which the formula evaluates to true. For example, the above formula is satisfiable, since setting $x_1$ to true, and $x_3$ to false (and $x_2$ and $x_4$ to any value, e.g., setting both to be true) causes the formula to evaluate to true.  Put another way, (True,True,False,True) would be said to be a so-called satisfying assignment of that formula.

Perhaps the single most important set in modern complexity theory is the set ``satisfiability,'' defined as $\sat = \{F \condition F$ is a satisfiable Boolean formula$\}$. The above formula $F(x_1,x_2, x_3,x_4) = (x_1 \land \overline{x_3}) \lor (x_2 \land \neg (x_3 \land x_1))$ is a member of $\sat$. But, for example, the formula $F(x_1,x_2) = x_1 \land \overline{x_1} \land x_2$ is not a member of $\sat$ (i.e., is not satisfiable)

One of the most important concepts in dealing with algorithms about Boolean formulas is the notion of self-reducibility. This is a type of divide-and-conquer approach to seeing whether a Boolean formula is satisfiable.  In particular, assuming $m \geq 1$, self-reducibility is the observation that (where to denote that $x_i$ is assigned a given truth value we will replace it in the call list with the assigned value): $F(x_1,x_2, \dots,x_m) \in \sat$ if and only if ($F(x_1,x_2, \dots,\falseit) \in \sat \lor F(x_1,x_2, \dots,\trueit) \in \sat$).  Please take a moment to chat with your groupmates to make sure you see why both the if and the only-if directions hold.

Note that we can think of those three formulas as a shallow tree, with $F(x_1,x_2, \dots,x_m)$ as the parent, having $F(x_1,x_2, \dots,\falseit)$ as its left child and $F(x_1,x_2, \dots,\trueit)$ as its right child:\nopagebreak
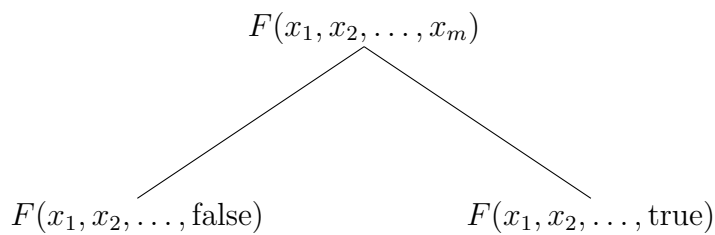
\begin{figure*}[h]
  \ctikzfig{fig-2-levels}
\end{figure*}

But note that we can continue this.  We can for each of those children give \emph{it} a left child with $x_{m-1}$ set to false
and a right child with $x_{m-1}$ set to true.  (So we have 4 nodes at that level, each with the final two variables fixed to a value.)  And we can continue that until all $m$ of the variables are
assigned.  At its bottom level, this tree has $2^m$ nodes, each with a complete assignment to the variables.

Note that it follows from inductive use of the self-reducibility rule---$F(x_1,x_2, \dots,x_m) \in \sat$ if and only if ($F(x_1,x_2, \dots,\falseit) \in \sat \lor F(x_1,x_2, \dots,\trueit) \in \sat$)---that $F(x_1,x_2, \dots,x_m) \in \sat$ if and only if one or more of those $2^m$ leaf nodes is an expression that evaluates to true.
(Though this fact is so directly obvious that one might wonder what the big deal is.)
This tree is known as a self-reducibility tree.  The size of the full tree is, as just mentioned, exponential in the number of variables---and it just reflects the fact that one can brute-force check all possible assignments to see if they are satisfied.

But the big deal about self-reducibility trees is the possibility that
one can find a way to ignore almost all of the tree, and explore just a polynomially large portion of the exponential-sized tree---and yet still be able to definitively determine whether the initial formula $F$ is satisfiable.

If we could do that in general, we would have resolved what is known as the ``P versus NP'' problem, would be awarded a million-dollar prize by the Clay Foundation, and would surely win the Turing Award.

This assignment is ambitious, but not \emph{that} ambitious.

Rather, the line that Fortune was working within seeks to show that \emph{under some assumption}, one in fact can
get away with exploring just a polynomial-sized portion of the tree. (And so it follows that either the assumption is untrue or one has won a million dollars.)

The hypothesis/assumption Fortune was studying is that $\satbar$ polynomial-time mapping reduces to a sparse set. Let us explain the three concepts in that sentence, each of which is a very important notion in theoretical computer science.
\begin{enumerate}
\item $\satbar$ is the complement of $\sat$, so it is all unsatisfiable Boolean formulas (and also all strings that are syntactically malformed; but such malformed strings can be recognized in polynomial time and so for the purpose of this assignment, let us henceforth completely ignore them, e.g., to avoid technical problems, let us simply assume that every malformed string is interpreted as being the well-formed---but not satisfiable---formula consisting just of the atom ``false'').
\item A set is sparse if it has at most polynomially many strings up to each given length. Formally, a set $S$ is sparse if there exists a polynomial $p$ such that, for each natural number $n$, the number of strings in $S$ of length less than or equal to $n$ is at most $p(n)$.  So the set $\{0,1,00,11,000,111,0000,1111,\dots\}$ is sparse, as it contains $2n$ strings of length at most $n$. But the set of all strings over the alphabet $\{0,1\}$ is not sparse, since it has $2^{n+1}-1$ strings of length at most $n$, and that exponential growth rate exceeds any polynomial.
\item A set $B$ polynomial-time mapping reduces to a set $C$ exactly if there exists a polynomial-time computable function $g$ such that, for each string $x$, it holds that $x \in B \iff g(x)\in C$. That is, $g$ quickly transforms a membership question about $A$ into a membership question about $B$. As an example, if we think for a moment about nonnegative integers rather than strings, we have that $N$ is even if and only $N+1$ is odd, so the function $f(N)=N+1$ mapping reduces from the set of even nonnegative integers to the set of odd nonnegative integers.

  For the rest of this assignment document, and also in the solutions document, we will by ``mapping reduces'' and ``mapping reduction'' always mean, respectively, ``polynomial-time mapping reduces'' and ``polynomial-time mapping reduction.''

\end{enumerate}

So Fortune's goal was to, under the assumption that there is a sparse set
that $\satbar$ mapping reduces
to, give a polynomial-time algorithm for $\sat$.

Yes, that was a lot of preparation. But we now are ready to give the general framework that had been established, before Fortune's work, for problems of this sort, and that Fortune of course knew of as he began his research, and he worked within it. You will also.

The framework is as follows. Our goal is to have to explore at most a polynomially large portion of the self-reducibility tree for whatever Boolean formula is given as our input, where our goal is to determine in polynomial time whether that formula belongs to $\satbar$ (or to determine in polynomial time whether that formula belongs to SAT; since a string is in SAT exactly if it is not in $\satbar$, clearly SAT is computable in polynomial time if and only if $\satbar$ is computable in polynomial time).

We want to do so by brutally trimming the tree. But if we build the whole tree first and then try trimming it, we are doomed from the start, since each self-reducibility tree's size is exponential in the number of variables in the formula (and so for many inputs is exponential in the size of the formula).

We are working under Fortune's assumption (hypothesis), so in our construction we may use $S$ as being a fixed sparse set to which $\satbar$ mapping reduces via the polynomial-time computable function $g$.  That is what $S$ and $g$ will refer to in the following.

So what the framework does is it \emph{trims the tree on the fly}.  It starts with a 1-node tree, namely, with the input formula, $F$.
And then the framework builds downward, level after level. For a given level that we have just finished with, for each node at that level the framework creates two child nodes (via splitting the next variable that has not already been assigned at the previous levels, with the left child assigning that variable to false and the right child assigning that variable to true). (Our notion of ``next'' counts down using the subscript of $x$. So the first split is on $x_m$, and the next is on $x_{m-1}$, and so on.)

That would eventually just build the whole, exponential-sized tree, so there must be more to the framework. And there is.
Each time we have just created a level, we exploit the fact that $\satbar$ mapping reduces to the sparse set $S$, via the polynomial-time computable function $g$. Namely, for each node at the level, we compute what $g$ outputs when its input is the formula that that node contains. The easiest way to think of this is that each tree node will in addition to the formula it is about, say $F'$, also contain an additional field---let us call it the node's S-label---that will contain $g(F')$. Utterly crucially, note that we have, due to the definition of mapping reductions, that $F' \in \satbar$ if and only if $g(F')\in S$.

Then, after having added to each node at that level that S-label field, the framework has one prune off (delete it from the tree) as many nodes of that level as one can (within polynomial time) consistent with ensuring that \emph{if at least one of the nodes at the level is about a satisfiable formula, then after we do our pruning at the level, we have left at that level at least one node that is about a satisfiable formula}.  (When we prune a node, it is gone, and so when we go to the next level, we won't build its two children at that level---we only build the children of those nodes that survive the pruning.)

Furthermore, the framework, as to doing that pruning, never does it based directly on looking at the formulas that the nodes at that level are capturing.  So even if a formula was ``$\falseit \land (\dots)$,'' the framework would not let one say ``this clearly is an unsatisfiable formula and so has no satisfying assignments and so we can prune this node.'' The reason the framework never does that is not that it might not sometimes allow useful pruning, but that it is a red herring---it does not seem to always lead one to a polynomial-time algorithm for the problem, and so is a distraction.

Let us look at this in action. So the input for the polynomial-time algorithm we are trying to build for SAT (or $\satbar$, but let us phrase it in terms of SAT since as argued above, either both or neither of SAT and $\satbar$ have polynomial-time algorithms) is some Boolean formula---one that has some number of variables (a finite number, but different inputs might have different numbers of variables).  Suppose a given such input formula has $m$ variables, and let us without loss of generality assume that the variables will always be named, in such a case, $x_1,\dots,x_m$.  So we can refer to the formula as $F = F(x_1,x_2, \dots,x_m)$.  What do we do? It is clear: We compute (in polynomial time) $g(F(x_1,x_2, \dots,x_m))$ and add it as the S-label of that root node.

So can we use whether $g(F(x_1,x_2, \dots,x_m))$ does or doesn't belong to $S$ to determine whether $F$ is, respectively, not in or in $\sat$ (that is not a typo---since $g$ maps from $\satbar$ to $S$, if $g(F)$ is in $S$ it means that $F$ is in $\satbar$, which is the same as saying that $F$ is not in $\sat$). The problem with that approach is that it doesn't work, because although $S$ is a sparse set, for all we know it could be some horribly uncomputable sparse set. Even with $g(F)$ in hand, we can't in general tell whether $g(F)\in S$.  So there is no obvious pruning we can do at this root level.  No problem. We move on by creating the two children of the root node (by assigning the root node's $x_m$ variable to false for the left child and to true for the right child), and we compute and add to each of those child nodes their S-labels (by applying $g$ to their formula).  And then we try at this second level the pruning method that you are about to (as Question~1) attempt to discover.  And then all nodes that didn't get pruned off at that level get split,
and we thus make the next level (where the final two variables of each node are fixed to true or false), and we assign its S-labels, and we do pruning there. And so we proceed, level after level.

For now, assume that the pruning is so excellent that the tree never becomes more than polynomially wide at any level. Then eventually we reach the final level---the one where all $m$ variables are assigned.  Each node we make at that bottom level is a complete assignment to $F$, so it is easy to evaluate $F$ under that assignment, and see whether the assignment does or does not satisfy $F$.  Note that (under the above temporary assumption about the tree not getting too wide, if that assumption is true---we'll have more on that later, as you'll need to do Question-2 magic to make that assumption true!)~we reach that with it and all earlier levels being at most polynomially wide. And by the rules of the pruning, we know that if there is \emph{any} satisfying assignment to $F$, then one of the polynomial number of assignments at this bottom node of the tree will satisfy $F$. So if any of them satisfies $F$, $F$ obviously is satisfiable. But if none of that polynomial number of assignments on that bottom level of our pruned tree satisfy $F$, then we know that neither those nor in fact any of the $2^m$ possible assignments satisfy $F$, even though we explored only a polynomially small part of the self-reducibility tree.

And now it is time for you to do the magic to bring this framework to life in this case, as Fortune did. You'll do so by, ideally, finding the answers to two questions.

\subparagraph{Question 1}

Basically, Question~1 is to find the pruning (aka trimming) rule to use at each level, right after your algorithms has filled in all that level's S-labels (and at the end of the section, after some more preparation work, our actual Question~1 will be stated). Remember, your goal is ensure that you prune off (delete it from the tree) as many nodes as you can (within polynomial time) consistent with ensuring that \emph{if at least one of the nodes at the level is about a satisfiable formula, then after we do our pruning at the level, we have left at that level at least one node that is about a satisfiable formula}.  (And we are going to ignore the red herring of using, other than via the S-label generated from it, the content of the formula at the node---basically since testing satisfiability by brute force currently cannot in general be done in less than exponential time.)

So please go to it! For example, since as CS people we often start with simple special cases, consider the moment when your algorithm has added the two children of $F$ (the left child has $x_m$ set to false, and the right child has $x_m$ set to true), and you have added to each of those two nodes the S-label value, which for the left node will be $g(F(x_1,x_2, \dots,\falserm))$ and for the right node will be $g(F(x_1,x_2, \dots,\truerm))$.

Can you see any case where you can safely delete one of the two nodes in a way that meets that italic requirement two paragraphs before this one?
If so, you are well on your way to formulating a pruning rule. (If not, please think some more about this until your group does find an answer.)

Next, think more generally. Suppose we're at the 6th level of our tree, i.e., the level with the final 5 variables---$x_{m-4}$, $x_{m-3}$, $x_{m-2}$, $x_{m-1}$, and $x_m$---instantiated.  So we may have as many as $2^5=32$ nodes as we start at this level. But suppose that due to earlier levels' prunings, we happen to start the level with 28 nodes. And of course, for each of those 28 nodes, your algorithm has used $g$ to compute its S-label. What pruning rule should we use for more general cases, such as this one?

\emph{(Question 1) Can you generalize what you (with luck, or actually, with skill and insight) did in the ``Can you see any case'' paragraph two paragraphs before this one to as strongly as possible (within polynomial time) trim nodes at a (general) level of the tree, still consistent with our goal of making sure that if at least one of the nodes at the level was about a satisfiable formula, then at least one of the nodes at that level left after your pruning will be satisfiable?}

Please put your group's answer---its pruning rule---down, as the answer to Question~1, on your group's joint hand-in answers.

If you've done that well, you've found one of the two key tree-size-containment actions of Fortune's algorithm. Congratulations!  (Historical tidbit: Fortune in fact did not discover this type of pruning himself, so I've asked you to do some work that Fortune didn't have to create himself; rather, this pruning rule was noticed and used in an earlier paper by Piotr Berman that Fortune was building on.  Berman's paper had appeared in one of the big four CS-theory conferences.)

\subparagraph{Question 2}

So it looks like the above gives a polynomial-time algorithm for SAT\@.
Are we done?

No; there is one flaw. The above discussion assumed that there is some polynomial that will bound how wide the tree gets, if one uses the Question~1 pruning rule (note: in Question~2 we are assuming that you got the same excellent pruning procedure that Fortune, drawing it from Berman's earlier work, used; my guess is that you likely did get it).

The problem with this assumption that the tree will get at most polynomially wide under the Question-1 pruning is that the assumption isn't even true!

To see that, consider the case where
the clever pruning never prunes anything. And even Fortune's Question-1 pruning rule---which I'm assuming is your pruning also---can have that happen.  So at the $(i+1)$'st level, where the last $i$ of the variables ($x_{m+1-i}$ through $x_m$) are set to false/true, we'd have $2^i$ nodes both before and after the pruning.  So nothing at all would be pruned from the tree, ever, and we'd end up with the entire exponential-sized self-reducibility tree, and so our algorithm would run in exponential time.  Not good!

To handle this, we need to get to the deepest magic of Fortune's work---an absolutely sparklingly clever insight.

Berman's precursor of Fortune's work actually had something not mentioned above: it had a bail-out case...~a situation when one can just stop abruptly at some level when one is perhaps nowhere near the bottom of the entire tree, and can declare that one knows for sure what the root is as to being in or not in SAT\@.  (To be fair to Fortune, in Berman's work that bail-out case is so closely tied to the pruning that Berman's bail-out case is a rather different animal than Fortune's bail-out case.)

You now need to make the huge leap Fortune made. To get to that, let us consider first the extreme case mentioned above, where Fortune's (which we assume is also your) Question-1 pruning rule on some particular input turns out to never prune any nodes, and so each level has twice as many nodes as the one before it: level after level, doubling in number of nodes with each new level.

Fortune's core insight was to argue that if that happens (or even the milder case holds that we prune but ``not enough,'' and so the tree grows ``too'' broad at some level), then that fact itself means we can stop right there, forget about doing anything else, and can declare, correctly, that the tree's root formula, $F = F(x_1,x_2, \dots,x_m)$, is satisfiable.

\emph{(Question 2) Can you give (i.e., please do give) a particular polynomial in $|F|$, call it $q$, and for that $q$ prove that if any level of your tree, after the level is pruned using your Question~1 pruning rule, has more than $q(|F|)$ nodes left, then that fact itself immediately implies that $F \in \sat$?}

By $|F|$ we mean the length (in characters) of the formula $F$ (technically, in $F$'s encoding, but let us not worry about encoding details in this assignment).  Since each variable certainly takes at least one character to include in a formula, clearly formula $F$ has no more than $|F|$ variables in it. So
we have that $m \leq |F|$.  Also, you are allowed to assume that for each node, say $F'$, in the entire self-reducibility tree of $F$, it holds that $|F'| \leq |F|$.\footnote{Why? Well, $|F|=|F|$, so this holds at the root, trivially. And at each level below the root, the formula is simply $F$ but with one or more variables assigned to be fixed to be true or to be false. But we may assume that when doing that assignment, we do the trivial simplifications that those logical base values allow, and doing so indeed ensures that fixing values never increases a formula's length. For example, as to the value true, note that the expression ``$\truerm \lor \mbox{\rm{}FOO}$'' (or the expression ``$\mbox{\rm{}FOO} \lor \truerm$) can be simplified to ``true''; the expression ``$\truerm \land \mbox{\rm{}FOO}$'' (or the expression ``$\mbox{\rm{}FOO} \land \truerm$) can be simplified to ``FOO''; and $\neg$true (or $\overline{\truerm}$) can be simplified to false, which might actually seem longer but in our coding of formulas we assume that both true and false are atoms in our logic and are coded as being of the same length as each other. Similar rules can be easily written for false.  Note that in those simplifications, we are assuming that careful attention is paid to
  precedence of operators when deciding when it is legal to apply a simplification.}

Note that your polynomial $q$ will likely have to be stated in terms of some of the other polynomials that are involved in the problem statement.  So $q$ can itself be stated in terms of for example (a)~the polynomial $p$ that bounds the growth of the sparse set $S$, i.e., the polynomial $p$ meeting the role of the $p$ from the definition earlier in this assignment of what it means for $S$ to be a sparse set, and (b)~the polynomial, call it $r$, that bounds the running time of the mapping reduction $g$ from $\satbar$ to $S$ in terms of the size of the input to the reduction. (Remember, we are working under the hypothesis that there is a sparse set $S$ such that $\satbar$ mapping reduces to $S$; so it is totally fair game to in our pruning speak of the density polynomial $q$ and the running-time polynomial $r$ as symbolic items we can use, even if we don't know what particular polynomial, e.g., $3n^8-n^2+4$, that symbol is allegedly realized by).

To be fair, as this is something that you might not yet have seen, I should mention to you, since this will be relevant in your solution, that in the machine model (the function-computing version of so-called Turing machines) that is used when studying the action of reductions (i.e., programs that are computing and output the value of a function), it holds that no program can, in a single step, output more than one character. For example, if the reduction $g$ runs for 17 steps on some input $w$, then $g(w)$ will have at most 17 characters.

Again, we have a lot of pieces in motion here, and so Question~2 is a big ask. However, to get a handle on it, think hard about the extreme case mentioned earlier in which our Question-1-type pruning never removes any nodes, so our tree's width doubles from each level to the next.
Think about your Question-1 pruning algorithm, and think about what it means about the S-labels of the nodes at a given level if no nodes were removed.  Do you see what that says about their labels?  Yet we're mapping each of the formulas at that level to $S$, which is a sparse set, and we know by the definition of mapping reductions that that means that each node at that level will belong to $\satbar$ exactly if its S-label belongs to $S$.  \emph{Can you see how this extreme case eventually will cause a huge tension that allows you to know for sure that the tree root, $F$, is satisfiable? And once you see that, can you pull back and extract the more general behavior at work here---namely, can you cleanly state, in terms of the polynomials $p$ and $r$, a polynomial $q$ such that if any level of the tree, after pruning, has more than $q(|F|)$ nodes, then we can immediately and correctly declare that $F \in \sat$?}

Please put your answer to Question~2 in your group's joint hand-in answers.

If you got this right, then
warmest congratulations, as your group has itself proved Fortune's result: If $\satbar$ mapping reduces to a sparse set, then $\sat$ is polynomial-time computable. It is great if your group got Question~1 right, and it is even better if your group also got Question~2 right.

After all, Fortune's result was an important advance in complexity and appeared in an utterly top-quality venue. In fact, Fortune's result was the precursor to the attainment, some years later by Steve Mahaney, of a long-elusive holy grail of complexity theory, namely, proving that if $\sat$ mapping reduces to a sparse set, then $\sat$ is polynomial-time computable.

\subparagraph{A Hint If You Are Stuck on Question~2} Please do your best, with your groupmates, to solve Question~2 with just the above information/framework. However, if you got Question~1 and have put in a good number of hours on Question~2 and still are just flat-out stuck, I've included as Appendix~A a hint. Feel free to look at it if you're sure that without it you won't be able to make progress. (I've put a page break just before Appendix~A, so that you don't unintentionally see the hint while reading this assignment.)

\subparagraph{Postscript: What You've Been Doing, in the Lingo}

This assignment was not assuming that you know the notions of co-nondeterministic and nondeterministic polynomial-time Turing machines, or their corresponding complexity classes, which are known as coNP and NP\@.  However, if you got Questions 1 and 2 correct, then you have proven a result about those, namely, Fortune's result.  In particular, basically because $\satbar$ plays a very central role in the study of coNP, in the (somewhat off-putting) jargon of complexity theory you in effect have actually proved the result that if there is a coNP-$\leq_{m}^{p}$-complete (or even a coNP-$\leq_{m}^{p}$-hard) sparse set, then $\textrm{P} = \textrm{NP}$; and you have, equivalently and still in the jargon of complexity theory, proved that if there is an NP-$\leq_{m}^{p}$-complete (or even a NP-$\leq_{m}^{p}$-hard) co-sparse set, then $\textrm{P} = \textrm{NP}$.

\clearpage

\appendix

\section*{Appendix A: A Hint on Question~2 (note: please do not read this unless you are outright stuck, even after putting in a lot of time, on Question~2)}

As noted in the main body of this assignment, please do your best, with your groupmates, to solve Question~2 with just the information/framework in the main body of the assignment. However, if you got Question~1 and have put in a good number of hours on Question~2 and just are flat-out stuck, below is a hint.  Please feel free to look at it if you're sure that without it you won't be able to make progress.

And here is the hint.
Suppose that in our course, at some class session at which 22 or more students are attending class, I say, ``I've just assigned the course grades, and at most 21 of you in this room will get a grade of A-minus or higher.''  What conclusion can you draw? And can you go back to Question~2 and take inspiration from the insight that conclusion holds?

  \clearpage
  \fancyfoot[C]{Solutions Preface Page, Just for the Instructor}
  
  \section*{Appendix B: Solutions (note to instructors: please do not include this in what you hand out to students as the assignment; rather, in addition to whatever review of answers is done in class, it is intended to provide a resource for students after their groups have tried and submitted their own assignment solutions)}

  The solutions start on the following page, and are numbered to start at page 1, so that the numbering makes sense to the students if these are given to them as a sample answer set.

  The solution set also provides---in case students wish to look at the original papers or to read about further progress in this line of work---a set of literature pointers along with a complete bibliographic reference for each mentioned paper.
  
  \clearpage
  \setcounter{page}{1} \fancyfoot[C]{Solutions Page \#\thepage}

  \sloppy
  \begin{center}
    {\large\em
      Solutions: {Fortune's Bounty: %
        Taming Complexity by Trimming Trees}}
  \end{center}

  {\bf Solution to Question 1:} Suppose that at some level, right after we create that level from the one before it and add all its S-labels, there exist one or more strings $y$ such that $y$ is the S-label of two or more nodes at this level.

  Let us think about what that means.  Let $y$ be any such string.  And so there are two, or three, or etc.\ nodes at that level of the tree that have the label $y$. Let $N_a$ be one such node and let $N_b$ be a different such node, both at that level of the tree.

  Since $N_a$ is mapped by $g$ to $y$, by the definition of mapping reductions we have that the formula $N_a$ is about belongs to $\satbar$ if and only if $y \in S$.

  Since $N_b$ is mapped by $g$ to $y$, by the definition of mapping reductions we have that the formula $N_b$ is about belongs to $\satbar$ if and only if $y \in S$.

  Though we offhand have no idea whether $y$ is or is not a member of $S$, by the transitivity of ``if and only if'' the above two statements let us conclude that the formula that $N_a$ is about belongs to $\satbar$ if and only if the formula that $N_b$ is about belongs to $\satbar$.  But that is equivalent to the statement that the formula that $N_a$ is about belongs to $\sat$ if and only if the formula that $N_b$ is about belongs to $\sat$.

  Ker\emph{POW}!  Why on Earth would we want to leave both $N_a$ and $N_b$ in our tree? After all, they stand or fall together as to being about a satisfiable formula.  So if we remove one of them, even if it was satisfiable, the other one would still be in and would be satisfiable also.  So if we delete one of them, that trimming certainly is consistent with our goal of ensuring that \emph{if at least one of the nodes at the level is about a satisfiable formula, then after we do our pruning at the level, we have left at that level at least one node that is about a satisfiable formula}.

  The above was just looking at \emph{two} items that share the S-label $y$.  If $y$ is the S-label of $j$ nodes, $j \geq 2$, at the given level, then we of course will want to keep in the tree \emph{one} of those $j$ nodes and trim from the tree the remaining $j-1$ of those nodes. (Which one of the $j$ we choose to keep doesn't matter, though if one wants to be specific, let us say we'll keep the leftmost node at that level that has $y$ as its S-label.)

  And of course, if more than one S-label occurs in two or more nodes at our current level, we'll do the above for each such S-label.

  In summary, our pruning rule (which was Fortune's rule, who himself brought it forward from the earlier work of Berman) is: After we enter a level and compute each of its nodes' S-labels, for each string $y$ that is the S-label of at least two nodes at that level, trim from the tree all nodes at that level with S-label $y$ except the leftmost one.

  {\bf Solution to Question 2:} The Appendix A hint, which you might not have seen if you tackled Question~2 without looking at that optional hint, was the following: ``Suppose that in our course, at some class session at which 22 or more students are attending class, I say, `I've just assigned the course grades, and at most 21 of you in this room will get a grade of A-minus or higher.'  What conclusion can you draw? And can you go back to Question~2 and take inspiration from the insight that conclusion holds?''

  The conclusion the hint was hoping for was that at least one of the students in the room would not get a grade of A-minus or higher.  Because if there are just 21 of those to go around, and we have 22 or more distinct students in the room on that day, at least one of them does not belong to the set of students who will be getting an A- or higher.

  Let's
  see how this applies to our case.  First, as mentioned in the problem statement, we may assume that $F$ itself, i.e., $F(x_1,x_2, \dots,x_m)$, has the property that every node we explore in the tree is about a formula whose length is no greater than that of $F$. That is, for each node $F'$ in the tree, $|F'| \leq |F|$.

  Okay.  So what can we state as an upper bound on how long the strings are that can be S-labels of nodes in the tree we are building?  Keeping in mind the previous paragraph, it is clear that $r(|F|)$ is such an upper bound, where, recall, $r$ is the polynomial bounding the running time of the mapping reduction, $g$, which maps from $\satbar$ to $S$.  After all, if a program runs for at most $r(|F|)$ steps, it certainly cannot output a string of length longer than $r(|F|)$, since on each step a program---at least in the standard machine model used in theoretical computer science for reductions, which is known as a (function-computing) Turing machine---can output at most one character.  (Since you might not have known this about the machine model, the problem statement mentioned that you could assume that in the model used in such complexity-theoretic studies, no function-computing machine can output more than one character per step.)

  And how many strings are there in $S$ of length less than or equal to that longest length that can be reached in things mapped to from nodes of our tree? The definition of sparseness gives us the answer to that.  At most $p(r(|F|))$ strings in $S$ are of length less than or equal to $r(|F|)$.  And $p(r(|F|))$ is certainly a polynomial itself, as the set of polynomials is clearly closed under composition.

  Think of $p(r(|F|))$ as the number of students who are going to get an A-minus or higher!

  Now, suppose at some level after pruning as per the rule developed in the answer to Question~1, we have $1+ p(r(|F|))$ or more nodes remaining.  Then at least one of them didn't get an A-minus or higher!

  In particular, since after pruning no two nodes on our current level have the same S-label, if we have $1+ p(r(|F|))$ or more nodes left on the level after pruning, then we have $1+ p(r(|F|))$ or more distinct S-labels, each of length at most $r(|F|)$.  \emph{But there are only at most $p(r(|F|))$ strings in $S$ of length at most $r(|F|)$.  So at least one of the nodes left after pruning has an S-label \textbf{that does not belong to \boldmath$S$}.}  Since $g$ was a mapping reduction from $\satbar$ to $S$, we know that any node whose S-label maps to a string that does not belong to $S$ is a node whose formula is satisfiable!  So at least one of the $1+ p(r(|F|))$ or more nodes at this level is satisfiable.  And since it is a partial instantiation of $F$'s variables, that immediately gives us that $F$ is satisfiable. So a correct choice of the polynomial $q$ that Question~2 was asking you to create would be $q(n) = p(r(n))$, since if there is even one more node than that left at a level after pruning, then we know the root formula $F$ is satisfiable.  Of course, that is not the only possible answer, e.g., $q(n) = 2026 + 42p(r(n))$ would also be a correct answer.

  But, again, if at some level, after pruning, we have $1+ p(r(|F|))$ or more nodes remaining, we are done, can stop the entire process, and can declare that $F$ is satisfiable! And that is Fortune's
  most magical contribution.  To see how subtle Fortune's insight is, consider this question: Which of the $1+ p(r(|F|))$ or more nodes at this level is one that we can (in polynomial time) point to and state that it is definitely satisfiable?  The answer is: We do not know. From the fact that a level of the tree, after pruning, had at least $1+ p(r(|F|))$ nodes left, we correctly concluded that at least one of those nodes was about a satisfiable formula...~but the algorithm is not able to necessarily point to one specific node and say with certainty that the formula associated with it is satisfiable (and beyond that, the algorithm is nowhere close to being able to necessarily give to a complete assignment to all variables that is a satisfying assignment). The algorithm, if it hits this case, knows that the root is satisfiable, but does so without having obtained a specific satisfying assignment for the root.  What Fortune did is really quite subtle and unusual.

  Finally, let us mention why Fortune's algorithm (using the answers/approach in these answers to Questions 1 and 2) runs in polynomial time.  If, after our Question-1-type pruning, any level had at least $1+ p(r(|F|))$ nodes, then we'd have stopped right there. And so, except if we have that kind of early cutoff, each level ends with at most $ p(r(|F|))$ nodes.
  And of course, if $F$ has $m$ variables, the tree has at most $m+1$ levels. So we have at most $m+1$ levels, each with at most (though just briefly, before pruning) $2p(r(|F|))$ nodes. So the whole tree is polynomial in size, relative to $|F|$.  Also, all the actions we take---such as finding repeated labels at a given level, and pruning all but the first of nodes at a level with the same S-label, and calculating S-labels, and if we reach the bottom level with all variables substituted checking whether one of the (necessarily complete)~variable assignments there is a satisfying assignment---can clearly be done in polynomial time.  Therefore, (as always) under the hypothesis that $\satbar$ mapping reduces to a sparse set, the entire process mentioned runs in polynomial time. And the process does determine whether or not $F$ belongs to $\sat$.

  So this establishes Fortune's result. Intuitively, it says that SAT is so hard that it $\satbar$ cannot mapping reduce to a sparse set...~unless in fact, contrary to the widespread belief of modern computer science, SAT is unconditionally in polynomial time even without any assumptions.

  \subparagraph{Literature Pointers} Fortune's article appeared as: A Note on Sparse Complete Sets, Steven Fortune, SIAM Journal on Computing, 1979~\cite{for:j:sparse}.

  We mention that coverage of Fortune's Theorem---stated in a form
  slightly different from, but equivalent to, the one used here---and
  a very detailed presentation of a proof of Fortune's Theorem can be
  found in the textbook, ``The Complexity Theory
  Companion''~\cite[Theorem~1.4]{hem-ogi:b:companion}. In the
  antepenultimate paragraph of our ``Solution to Question 2,'' we
  noted that the given algorithmic proof of
  Fortune's Theorem would, when the tree got
  ``bushier'' than a specific threshold, be able to safely and
  correctly declare that $F \in \sat$, yet that given polynomial-time
  algorithm would not necessarily be able to point to
  a satisfying assignment for $F$,
  and would not even necessarily be able to point to some specific one
  of the still-live nodes at the given level and state that it was
  capturing a prefix of a satisfying assignment for $F$, despite the
  fact that at least one of the nodes indeed would be
  capturing such a prefix. If the reader wants
  an extra challenge to tackle, here is one. It in fact is known that by
  changing the flavor of that polynomial-time algorithm to a different
  algorithm that also runs in polynomial-time, one can avoid this
  behavior and always obtain satisfying assignment
  for a given satisfiable formula (under the assumption of Fortune's
  Theorem). Finding that variant algorithm is given as an exercise to
  the reader in the above-cited textbook (as its ``Pause to
  Ponder~1.6''), and is something the reader might enjoy tackling. 
  So that the reader can find a solution to this if it proves
  a hard nut to crack,
  we mention that the textbook as a footnote within
  that Pause to Ponder gives the reader a give-away hint of what
  change will yield the variant algorithm.
  
  Berman's paper, which Fortune was building on, appeared as: Relationship Between Density and Deterministic Complexity of NP-Complete Languages, Piotr Berman, Fifth International Colloquium on Automata, Languages, and Programming (ICALP), 1978~\cite{ber:c:relate}

  Fortune's work was part of a long line of research, and indeed, even more than twenty years after Fortune's paper, advances continued to be made in that line, see, e.g.,~\cite{arv-koe-mun:c:sparse,arv-han-hem-koe-loz-mun-ogi-sch-sil-thi:b:sparse,gla:t:sparse,gla-hem:j:clarityII}. However, the single most notable work in that line came not long after Fortune's work, and showed that Fortune's result remains true even if one replaces $\satbar$ with $\sat$ in Fortune's theorem. That result was proven by Stephen Mahaney: If SAT mapping reduces to a sparse set, then SAT is polynomial-time computable. Mahaney's article appeared as: Sparse Complete Sets for {NP}: {S}olution of a Conjecture of {B}erman and {H}artmanis, Stephen Mahaney, Journal of Computer and System Sciences, 1982~\cite{mah:j:sparse-complete}.

  %
\newcommand{\etalchar}[1]{$^{#1}$}

\end{document}